\documentclass[10pt]{bmc_article}

\usepackage{cite} 
\usepackage{url}  
\usepackage{ifthen}  
\usepackage{multicol}   
\usepackage[utf8]{inputenc} 
\usepackage{graphicx}
\usepackage{caption}
\usepackage{rotating}
\usepackage{multirow}
\usepackage{booktabs}

\newboolean{publ}

\newenvironment{bmcformat}{\baselineskip20pt\sloppy\setboolean{publ}{false}}{\baselineskip20pt\sloppy}

\begin{document}
\begin{bmcformat}

\title{Predicting genome-wide DNA methylation using methylation marks, genomic position, and DNA regulatory elements}

\author{Weiwei Zhang$^1$%
               \and
                 Tim D Spector$^2$%
               \and
                 Panos Deloukas$^3$%
               \and
        Jordana T Bell$^{2,7}$%
       and
       Barbara E Engelhardt\correspondingauthor$^{4,5,6,7}$%
         \email{Barbara E Engelhardt\correspondingauthor - barbara.engelhardt@duke.edu}%
       }

\address{%
    \iid(1) Department of Molecular Genetics and Microbiology, Duke University, Durham, North Carolina, USA\\
    \iid(2) Department of Twin Research and Genetic Epidemiology, King's College London, London, UK. \\
    \iid(3) Wellcome Trust Sanger Institute, Hinxton CB10 1SA, UK. \\
    \iid(4) Department of Biostatistics \& Bioinformatics, Duke University, Durham, North Carolina, USA\\
    \iid(5) Department of Statistical Science, Duke University, Durham, North Carolina, USA\\
    \iid(6) Institute for Genome Sciences \& Policy, Duke University, Durham, North Carolina, USA\\
    \iid(7) Equal contributors
}%

\maketitle

\begin{abstract}

\subsection*{Background}
Recent assays for individual-specific genome-wide DNA methylation profiles have enabled epigenome-wide association studies to identify specific CpG sites associated with a phenotype. Computational prediction of CpG site-specific methylation levels is important, but current approaches tackle average methylation within a genomic locus and are often limited to specific genomic regions.
\subsection*{Results}
We characterize genome-wide DNA methylation patterns, and show that correlation among CpG sites decays rapidly, making predictions solely based on neighboring sites challenging. We built a random forest classifier to predict CpG site methylation levels using as features neighboring CpG site methylation levels and genomic distance, and co-localization with coding regions, CGIs, and regulatory elements from the ENCODE project, among others. Our approach achieves $91\% -94\%$ prediction accuracy of genome-wide methylation levels at single CpG site precision. The accuracy increases to $98\%$ when restricted to CpG sites within CGIs. 
Our classifier outperforms state-of-the-art methylation classifiers and identifies features that contribute to prediction accuracy: neighboring CpG site methylation status, CpG island status, co-localized DNase I hypersensitive sites, and specific transcription factor binding sites were found to be most predictive of methylation levels. 
\subsection*{Conclusions}
Our observations of DNA methylation patterns led us to develop a classifier to predict site-specific methylation levels that achieves the best DNA methylation predictive accuracy to date. Furthermore, our method identified genomic features that interact with DNA methylation, elucidating mechanisms involved in DNA methylation modification and regulation, and linking different epigenetic processes.
\end{abstract}

\ifthenelse{\boolean{publ}}{\begin{multicols}{2}}{}

\section*{Keywords}
DNA methylation, CpG island, shore, and shelf, random forest classifier, DNase I hypersensitive sites, transcription factor binding sites, EWAS

\bigskip

\section*{Background}

Epigenetics is the study of changes in gene expression or complex phenotype that are not associated with changes in DNA sequence and but inherited through cell division. Epigenetic markers often change within an individual over time and are cell type specific~\cite{Barrero2010, Scarano2005, Cedar2012}. Epigenetics has been shown to play a critical role in cell differentiation, development, and tumorigenesis~\cite{Kiefer2007,Tost2010}. DNA methylation is probably the best studied epigenetic modification of DNA, but our understanding of DNA methylation is still in its infancy. In vertebrates, DNA methylation occurs by adding a methyl group to the fifth carbon of the cytosine residue, mainly in the context of neighboring cytosine and guanine nucleotides in the genome (5-CG-3 dinucleotides or \emph{CpG sites}) mediated by DNA methyl-transferases (DNMTs)~\cite{Cedar1988, Jaenisch2003}. DNA methylation has been shown to play an important functional role in the cell, including involvement in DNA replication and gene transcription, with substantial downstream association with development, aging, and cancer~\cite{Barrero2010,Wolffe1999,Rivenbark2012,Das2004,Scarano2005,Cedar2012}.

CpG sites are underrepresented in the human genome relative to their expected frequency as a result of being a \emph{mutation hotspot}, where the deamination of methylated cytosines often changes CpG sites into TpG sites~\cite{Tost2010,Lienert2011}. Although CpG sites are mainly methylated across the mammalian genome~\cite{Jones2012}, there are distinct, mostly unmethylated CG-rich regions termed CpG islands (CGIs) that have a G+C content greater than $50\%$~\cite{Tost2010,Lienert2011,Law2010}. CGIs account for $1-2\%$ of the genome and are often located in promoters and exonic regions in mammalian genomes~\cite{Shen2007,Larsen1992}. Methylation patterns in CGIs that are in promoter regions, where most previous studies have focused attention, have recently been shown to differ from methylation patterns elsewhere, indicating a specific biological role for these promoter CGIs~\cite{Jones2012}. CGIs have been shown to co-localize with DNA regulatory elements such as transcription factor binding sites (TFBSs)~\cite{Brandeis1994,Macleod1994,Dickson2010,Teschendorff2009,Deaton2011,Choy2010,Gebhard2010,Stirzaker2004} and DNA binding insulator proteins, such as CTCF, which protect downstream DNA from upstream methylation activities~\cite{Valenzuela2006}. Across the genome, DNA methylation levels have been shown to be associated with gene regions, active chromatin marks~\cite{Weber2007,Meissner2008,Hawkins2010}, cis-acting DNA regulatory elements, and proximal sequence elements~\cite{Shen2007,Das2006}, giving hints about the processes that regulate methylation and how methylation may in turn impact cellular phenotypes.

The non-uniform distribution of CpG sites across the human genome and the important role of methylation in cellular processes imply that characterizing genome-wide DNA methylation patterns is necessary to better understand the regulatory mechanisms of this epigenetic phenomenon~\cite{Laird2010}. Recent advances in methylation-specific microarray and sequencing technologies have enabled the assay of DNA methylation patterns genome-wide and at single base-pair resolution~\cite{Laird2010}. 
The current gold standard to quantify single site DNA methylation levels across an individual's genome is whole genome bisulfite sequencing (WGBS), which quantifies DNA methylation levels at $\sim 26$ million (out of $28$ million total) CpG sites in the human genome~\cite{Laurent2010, Lister2011,Lister2009}. However, WGBS is prohibitively expensive for most current studies, is subject to conversion bias, and is difficult to perform in particular genomic regions~\cite{Laird2010}. Other sequencing methods include methylated DNA immunoprecipitation (MeDIP) sequencing, which is experimentally difficult and expensive, and reduced representation bisulfite sequencing (RRBS), which assays CpG sites in small regions of the genome ~\cite{Laird2010}.
As an alternative, methylation microarrays, and the Illumina HumanMethylation 450K Beadchip in particular, measure bisulphite treated DNA methylation levels at $\sim 482,000$ preselected CpG sites genome-wide~\cite{Sandoval2011}; however, these arrays assay less than $2\%$ of CpG sites, and this percentage is biased to gene regions and CGIs. Quantitative methods are needed to predict methylation status at unassayed sites and genomic regions.

In this study, we examined measurements of methylation levels in $100$ individuals using the Illumina 450K Beadchip~\cite{Bibikova2011}.
Within these methylation profiles, we examined the patterns and correlation structure of the CpG sites, with attention to characterizing methylation patterns in CGI regions. Using features that included neighboring CpG site methylation status, genomic location, local genomic features, and co-localized regulatory elements, we developed a random forest classifier to predict single CpG site methylation levels.  
Using this model, we were able to identify DNA regulatory elements that may interact with DNA methylation at specific CpG sites, providing hypotheses for experimental studies on mechanisms by which methylation is regulated or leads to biological changes or disease phenotypes.


\subsection*{Related work in DNA methylation prediction}

Methylation status is a delicate epigenomic feature to characterize, and even more challenging to predict, because assayed DNA methylation marks are (a) an average across the sampled cells, (b) cell type specific, (c) environmentally unstable, and (d) not well-correlated within a genomic locus~\cite{Scarano2005, Bell2011, Eckhardt2006}. It is not clear whether either predicted or measured methylation status at specific CpG sites generalize well across platforms, cell types, individuals, or genomic regions~\cite{Fernandez2011}. A number of methods to predict methylation status have been developed (Table~S1). Most of these methods assume that methylation status is encoded as a binary variable, e.g., a CpG site is either methylated or unmethylated in an individual~\cite{Bhasin2005, Bock2006, Das2006, Fang2006, Kim2007, Fan2008, Lu2010, Zheng2013}. These methods have often limited predictions to specific regions of the genome, such as CGIs~\cite{Bock2006,Kim2007,Fang2006,Fan2008,Previti2009,Zheng2013}. More broadly, all of these methods make predictions of average methylation status for windows of the genome instead of individual CpG sites. Our model is an attempt to develop a general method for predicting methylation levels at individual CpG sites by removing the simplifying assumptions of current methods without sacrificing prediction accuracy.

The relative success of these methods depends heavily on the prediction objectives. All of the studies that achieved prediction accuracy $\geq 90\%$~\cite{Bock2006,Fan2008,Previti2009,Zheng2013} predicted average methylation status within CGIs or DNA fragments within CGIs. Most of the CpG sites in CGIs are unmethylated across the genome~\cite{Jones2012} -- for example, $16\%$ of CpG sites in CGIs in cells from human brain were found to be methylated using a WGBS approach~\cite{Maunakea2010} -- so it is not surprising that classifiers limited to these regions perform well. Studies not limiting prediction to CGIs uniformly achieved lower accuracies, ranging from $75\%$ to $86\%$.
Many of these methods predicted binary methylation status; one study used categorical methylation status~\cite{Previti2009}. Another study predicted average methylation levels as a continuous variable~\cite{Zhou2012}, and the predictions achieved a maximum correlation coefficient of $0.82$ and a root mean square error (RMSE) of $0.20$; however, the study was limited to $\sim 400$ DNA fragments instead of a genome-wide analysis.

Across these methods, features that are used for DNA methylation prediction include: DNA composition (proximal DNA sequence patterns), predicted DNA structure (e.g., co-localized introns), repeat elements, TFBSs, evolutionary conservation (e.g., \emph{PhastCons}~\cite{Siepel2005}), number of SNPs, GC content, Alu elements, histone modification marks, and functional annotations of nearby genes. Several studies only using DNA composition features achieved prediction accuracies ranging from $75\%$ to $87\%$~\cite{Bhasin2005,Das2006,Kim2007,Lu2010,Zhou2012}. Bock \emph{et al.} used $\sim 700$ features including DNA composition, DNA structure, repeat elements, TFBSs, evolutionary conservation, and number of SNPs; Zheng \emph{et al.} included $\sim 300$ features including DNA composition, DNA structure, TFBSs, histone modification marks, and functional annotations of nearby genes. Two studies did not use any DNA composition features~\cite{Fan2008, Previti2009}. The relative contribution of each feature to prediction quality is not quantified well across these studies because of the different methods and prediction objectives.

The majority of these methods are based on support vector machine (SVM) classifiers~\cite{Bhasin2005,Bock2006,Das2006,Fang2006,Fan2008,Previti2009,Zhou2012,Zheng2013}.
General non-additive interactions between features are not encoded when using linear kernels, as most of these SVM-based classifiers used. If a more sophisticated kernel is used, such as a radial basis function kernel (RBF), within the SVM-based approach, the contribution of each feature to the prediction quality is not readily available.  Three studies included alternative classification frameworks: one found that a decision tree classifier achieved better performance than an SVM-based classifier~\cite{Previti2009}. Another study found that a naive Bayes classifier achieved the best prediction performance~\cite{Kim2007}. A third study used a word composition-based encoding method~\cite{Lu2010}.  In our study, we choose to use a random forest classifier because it encodes non-additive interactions between features and because the relative contribution of each feature to prediction quality is quantifiable.

Our method for predicting DNA methylation levels at CpG sites genome-wide differs from these current state-of-the-art classifiers in that it: (a) uses a genome-wide approach, (b) makes predictions at single CpG site resolution, (c) is based on a random forest classifier, (d) predicts methylation levels instead of methylation status, (e) incorporates a diverse set of predictive features, including regulatory marks from the ENCODE project, and
(f) allows the quantification of the contribution of each feature to prediction. We find that these differences substantially improve the performance of the classifier and also provide testable biological insights into how methylation regulates, or is regulated by, specific genomic and epigenomic processes.

\section*{Results}

\subsection*{Characterizing methylation patterns}

DNA methylation profiles were measured in whole blood samples from $100$ unrelated individuals by Illumina HumanMethylation 450K Beadchips at single CpG site resolution for $482,421$ CpG sites~\cite{Heyn2013}. 
Single CpG site methylation levels are quantified by $\beta$, the ratio of the methylated probe intensity and the sum of the methylated and unmethylated probe intensities, which ranges from 0 (unmethylated) to 1 (methylated). Within the single-site $\beta$ values across individuals, we controlled for probe chip position, sample age, and sample sex. After these data were preprocessed (see Materials and Methods), $394,354$ CpG sites remained across the $22$ autosomal chromosomes.

First we examined the distribution of DNA methylation levels, $\beta$, at CpG sites on autosomal chromosomes across all $100$ individuals (Figure~1A). The majority of CpG sites were either hypermethylated or hypomethylated, with $48.2\%$ of sites with $\beta > 0.7$ and $40.4\%$ of sites with $\beta < 0.3$. Using a cutoff of $0.5$, across the methylation profiles and individuals, $54.8\%$ of these CpG sites have a \emph{methylated status} ($\beta \geq 0.5$). 
Across the individuals, we observed distinct patterns of DNA methylation levels in different genomic regions (Figure~1B). Using CGIs labeled in the UCSC genome browser~\cite{Kent2002}, we defined \emph{CGI shores} as regions $0 - 2$ kb away from CGIs in both directions and \emph{CGI shelves} as regions $2 - 4$ kb away from CGIs in both directions~\cite{Bibikova2011}. We found that CpG sites in CGIs were hypomethylated ($81.2\%$ of sites with $\beta < 0.3$) and sites in non-CGIs were hypermethylated ($73.2\%$ of sites with $\beta > 0.7$), while CpG sites in CGI shore regions had variable methylation levels following a U-shape distribution ($39.0\%$ of sites with $\beta > 0.7$ and $46.2\%$ of sites with $\beta < 0.3$), and CpG sites in CGI shelf regions were hypermethylated ($78.2\%$ of sites with $\beta > 0.7$). These distinct patterns reflect highly context-specific DNA methylation levels genome-wide.

\begin{figure}[h]
\centerline{\includegraphics[width=4in]{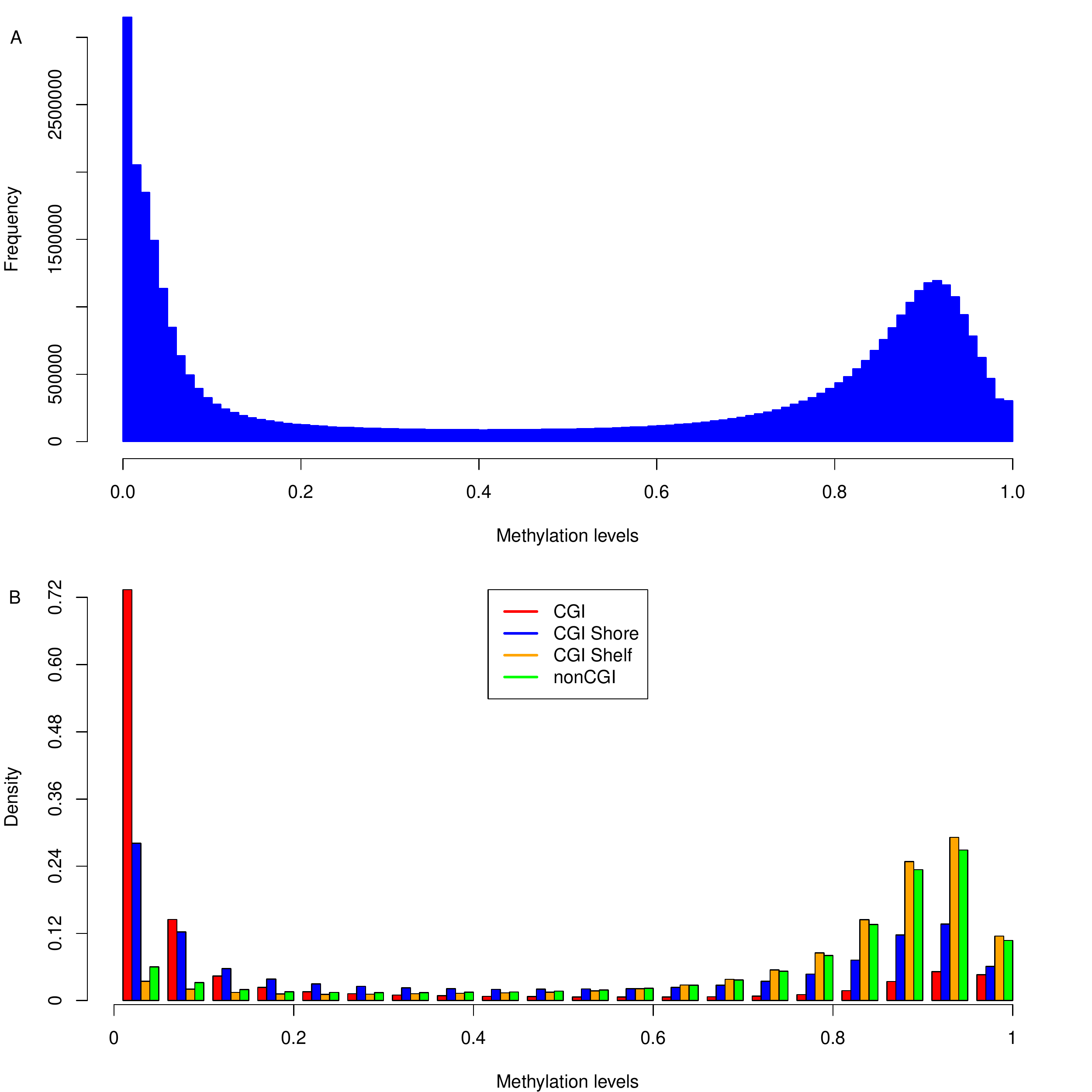}}
\caption{{\bf Distribution of DNA methylation levels at CpG sites across autosomal chromosomes.} Methylation levels across $100$ individuals at CpG sites assayed on the 450K array. Panel A: Distribution of DNA methylation values across all CpG sites. Panel B: Distribution of DNA methylation values for CpG sites within CGIs, CGI shores, CGI shelves, and non-CGI regions.}
\end{figure}

DNA methylation levels at nearby CpG sites have previously been found to be correlated (indicating possible co-methylation), particularly when CpG sites are within $1 -2$ kb from each other~\cite{Bell2011,Eckhardt2006}; these patterns are in contrast with the correlation among nearby genotypes due to linkage disequilibrium (LD) that often extends to large genomic regions from a few kilobases to $>1$ Mb~\cite{1000genomes}. 
We quantified the correlation of methylation levels ($\beta$) between neighboring pairs of CpG sites using the absolute value Pearson's correlation across individuals. We found that correlation of methylation levels between neighboring (or adjacent on the array) CpG sites decreased rapidly to approximately $0.4$ within $\sim 400$ bp, in contrast to sharp decays noted within $1-2$ kb in previous studies with sparser CpG coverage (Figure~2A)~\cite{Bell2011,Eckhardt2006}. 
We found the rate of decay in correlation to be highly dependent on genomic context; for example, for neighboring CpG sites in the same CGI shore and shelf region, correlation decreases continuously until it is well below what is expected (Figure~2A). Because of the over-representation of CpG sites near CGIs on the array, an increase in correlation can be observed as neighboring sites extend past the CGI shelf regions, where there is lower correlation with CGI methylation levels than we observe in the background.

\begin{figure}[h]
\centerline{\includegraphics[width=4in]{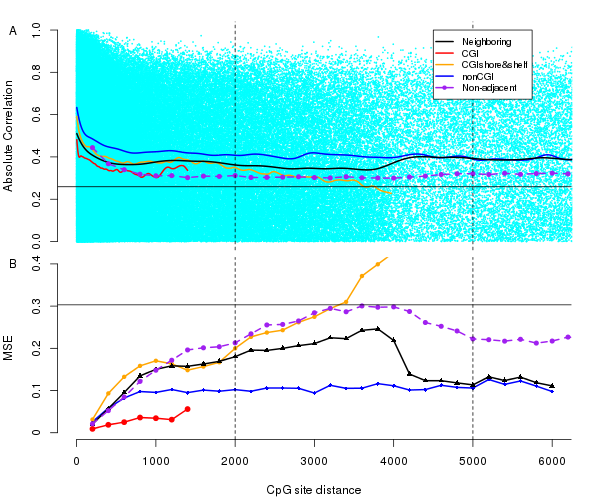}}
\caption{{\bf Correlation of methylation levels between neighboring CpG sites.}
      The x-axis represents the genomic distance in bases between the neighboring CpG sites, or assayed CpG sites that are adjacent in the genome. Different colors and points represent subsets of the CpG sites genome-wide, including pairs of CpG sites that are not adjacent in the genome but that are the specified distance apart (\emph{non-adjacent}). The CGI shore \& shelf CpG sites are truncated at $4000$ bp, which is the length of the CGI shore \& shelf regions.  The solid horizontal line represents the background (absolute value correlation or MSE) levels averaged from $10,000$ pairs of CpG sites from arbitrary chromosomes. Panel A: the absolute value of the correlation between neighboring sites across all individuals (y-axis). The lines represent cubic smoothing splines fitted to the correlation data. Panel B: the MSE was calculated for CpG sites (y-axis) for each pair of CpG sites within the genomic distance window.}
\end{figure}

To make this decay more precise, we contrasted the observed decay to the level of \emph{background correlation} ($0.259$), which is the average absolute value correlation between the methylation levels of pairs of randomly selected CpG sites across chromosomes (Figure~2A, Figure 3). We found substantial differences in correlation between neighboring CpG sites versus arbitrary pairs of CpG sites at identical distances, presumably because of the dense CpG tiling on the 450K array within CGI regions. Interestingly, the slope of the correlation decay plateaus after the CpG sites are approximately $400$ bp apart (both for neighbors and for arbitrary pairs of a specified distance), but the distribution of correlation between pairs of CpG sites is not substantially different from the distribution of background correlation even within $200$ Kb (Figure~S1A). 
While this certainly suggests that there are may be patterns of methylation that extend to large genomic regions, the pattern of extreme decay within approximately $400$ bp across the genome indicates that, in general, methylation may be biologically manipulated within very small genomic regions. Thus, neighboring CpG sites may only be useful for prediction when the sites are sampled at sufficiently high densities across the genome.

\begin{figure}[h]
\centerline{\includegraphics[width=4in]{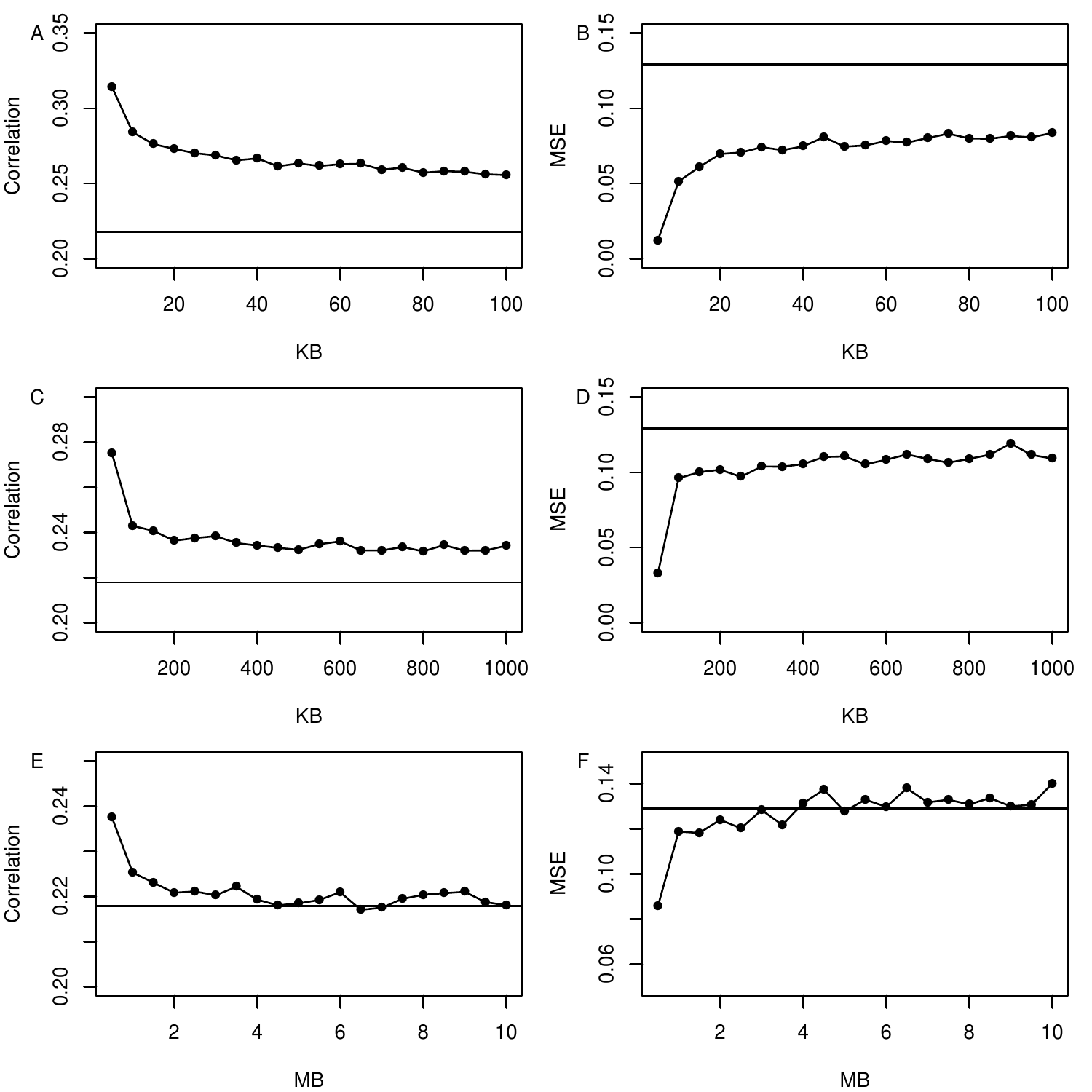}}
\caption{{\b Correlation and MSE of methylation values between arbitrary pairs of CpG sites}
     The x-axes represent the genomic distance between pairs of CpG sites; The left column shows the correlation of CpG sites within $100$ kb (Panel A), $1$ Mb (Panel C) and $10$ Mb (Panel E); the right column plots show the MSE patterns of CpG sites in relation to their genomic distances with distance range $100$ kb (Panel B), $1$ Mb (Panel D) and $10$ Mb (Panel F).
  The solid horizontal lines represent the background correlation or MSE level calculated from $10000$ pairs of CpG sites from a different chromosome.}
\end{figure}

We repeated these experiments using mean squared error (MSE) between CpG site levels to quantify patterns of decay of methylation within each individual, instead of across individuals as is measured with the correlation analyses (see Materials and Methods; Figure~2B, Figure 3). In general, the MSE trends echo the local patterns seen in the correlation analysis and also appear to be region specific. In CGI regions, the MSE of neighboring sites was low and increased slowly with genomic distance. In contrast, MSE in CGI shore and shelf regions increased rapidly to an MSE higher than background MSE ($0.30$), indicating that the edges of a single shore and shelf region are less predictive of each other than any two CpG sites at random. The individual-specific MSE between neighboring sites (Figure~S1B) shows a much higher deviation from background distribution of MSE at $200$ kb relative to correlation, indicating that the biological manipulation of methylation in larger genomic regions may be individual specific, such as being driven by genetic variants or environmental effects.

As we observed that methylation patterns at neighboring CpG sites depended heavily on genomic content, we further investigated methylation patterns within CGIs, CGI shores, and CGI shelves. Methylation levels at CGIs and CGI shelves were fairly constant genome-wide and across individuals -- CGIs are hypomethylated and CGI shelves are hypermethylated -- but CGI shores exhibit a reproducible but drastic pattern of change (Figure~4A). CpG sites in CGI shores have a monotone increasing pattern of methylation status from CGIs towards CGI shelves, and this pattern is symmetric in the CGI shores upstream and downstream of CGIs.
If we examine the MSE between pairs of CpG sites' methylation status in these regions, we find that MSE within the CGI and within the CGI shelves is low, consistent with the variance we observed within DNA methylation profiles in these regions (Figure~4B). Additionally, we find that the MSE between the CpG sites in the shelves appears to increase as the sites are further away from the CGI on the shelf, suggesting a circular dependency in methylation status between the ends of the shelf sequences. It is interesting that the CpG sites in the shore regions appear substantially more predictive of CpG sites in the shelf regions than those in the CGI regions, although this may indicate a less precise delineation of the shore and shelf regions relative to the CGI and CGI shore delineation.

\begin{figure}[h]
\centerline{\includegraphics[width=6in]{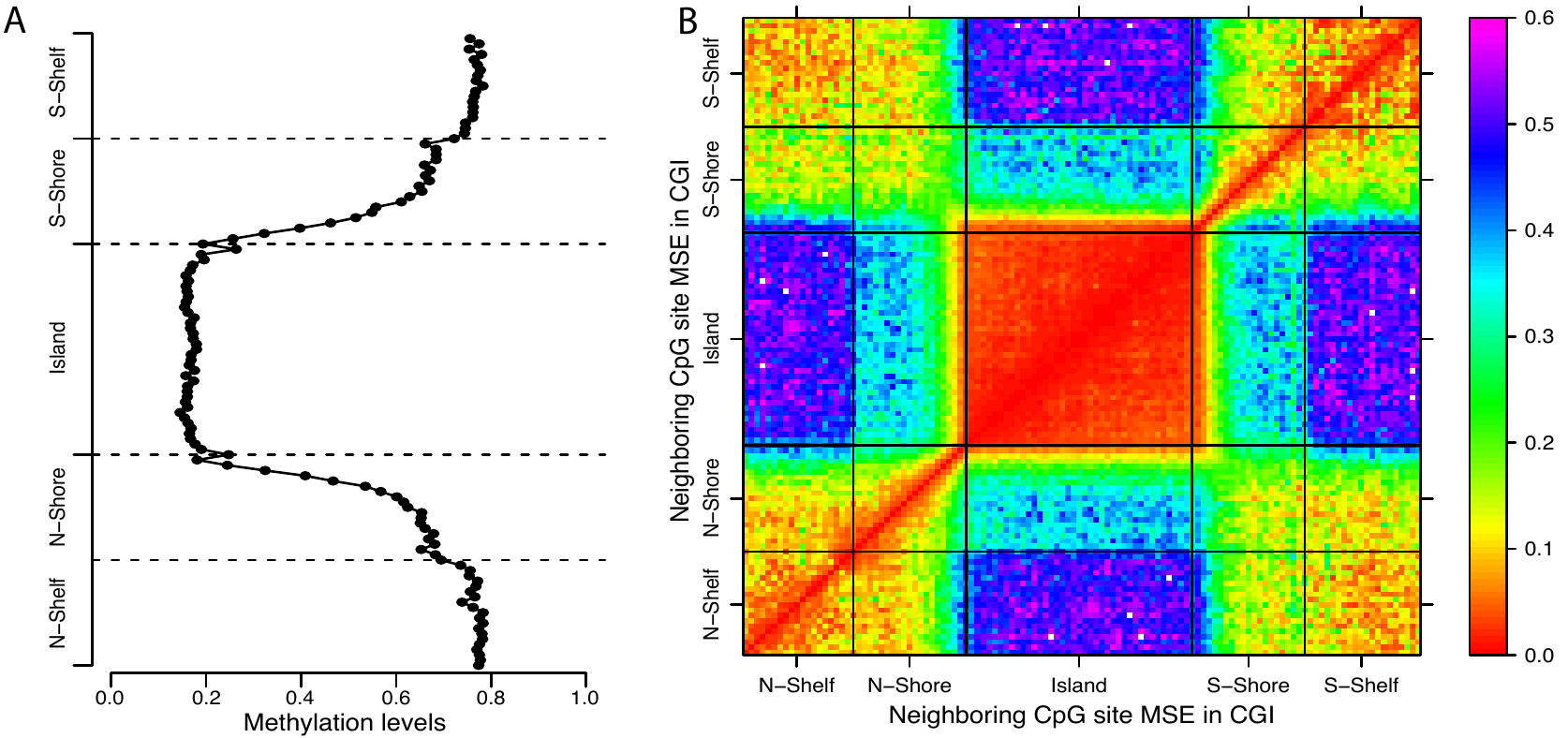}}
\caption{{\bf Figure 4: Methylation structure with respect to CpG islands status.} Since each CGI is a different length, each CGI was split into $40$ equal sized windows.
      Panel A: The points represents the mean $\beta$ in CGIs, CGI shores or CGI shelves across all sites in all individuals with a window size of $100$ bp.
      Panel B: Methylation levels of each CpG site in each CGI, CGI shore, and CGI shelf were compared with all the other sites in the same CGI region. X-axis and y-axis represent the genomic position of each CGI with a scale of 1:100, i.e. one unit in matrix represents $100$ bp distance. The MSE of each unit cell was calculated for all pairwise CpG sites with one site located in the relative scaled position on x-axis and the other one on y-axis, and then averaged over 100 individuals.}
\end{figure}

To quantify the amount of variation in DNA methylation explained by genomic context, we considered the correlation between genomic context and principle components (PCs) of methylation levels across all $100$ individuals (Figure 5). 
We found that many of the features derived from the CpG site's genomic context appear to be correlated with the first principal component (PC1). Methylation statuses of upstream and downstream neighboring CpG sites and a co-localized DNAse I hypersensitive (DHS) site are the most highly correlated features, both with Pearson's correlation around $0.57$ (Figures~5). Ten genomic features all have correlation $>0.5$ with PC1, including co-localized active TFBSs Elf1 (ETS-related transcription factor 1), MAZ (Myc-associated zinc finger protein), Mxi1 (MAX-interacting protein 1) and Runx3 (Runt-related transcription factor 3), suggesting that they may be useful in predicting DNA methylation status (Figure~S2). That said, the features themselves are well correlated; for example, active TFBS are highly enriched within DHS sites (correlation $r=0.66$)~\cite{Keene1981, Bernat2006}.

\begin{figure}[h]
\centerline{\includegraphics[width=7in]{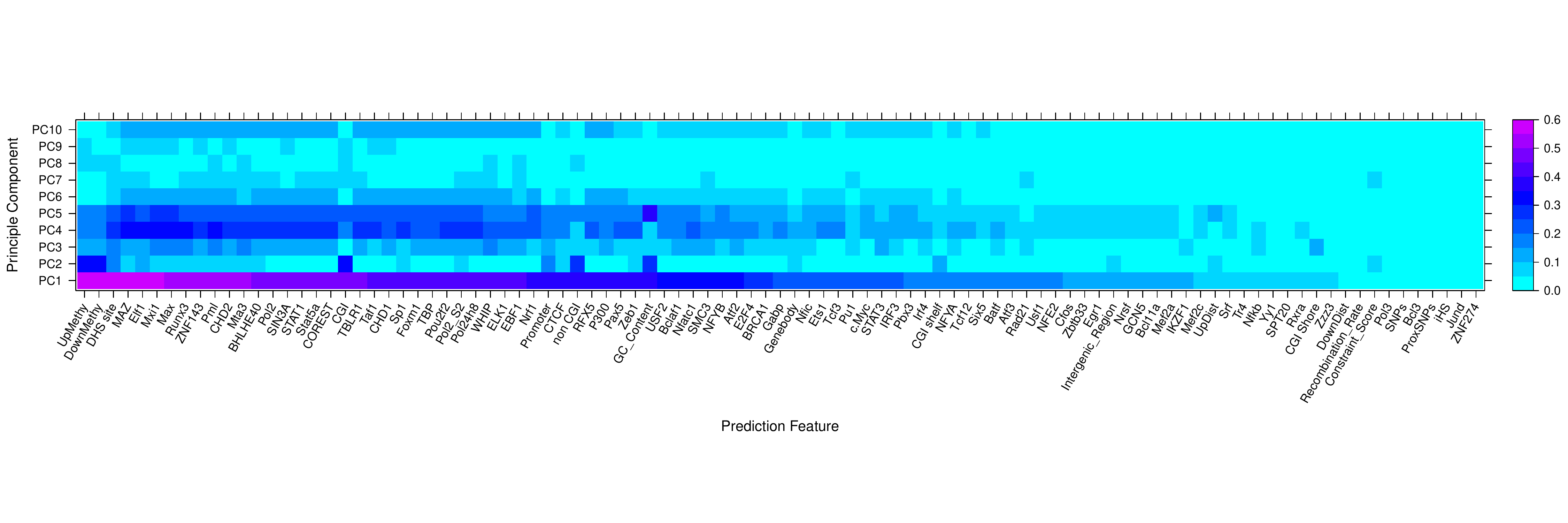}}
\caption{{\bf Correlation matrix of prediction features with first ten principle components of methylation levels.}
      PCA is performed on methylation levels for $37,865$ CpG sites. The correlation with each of the features was calculated between the first ten methylation PCs (y-axis) and all features (x-axis).}
\end{figure}

\subsection*{Binary methylation status prediction}

These observations about patterns of DNA methylation suggest that correlation in DNA methylation is local and dependent on genomic context. Thus, prediction of DNA methylation status based only on methylation levels at neighboring CpG sites may not perform well, especially in sparsely assayed regions of the genome. Using prediction features including neighboring CpG site methylation levels and a number of features characterizing genomic context, we built a classifier to predict DNA methylation status, where status indicates no methylation (0) or complete methylation (1) at a CpG site. There were $378,677$ sites with neighboring CpG sites at arbitrary distances used in these analysis per individual.

The $97$ features that we used for DNA methylation status prediction fall into four different classes (Table S4): 
\begin{itemize}
\item \emph{neighbors}: one upstream and one downstream neighboring (CpG sites assayed on the array and adjacent in the genome) CpG sites' genomic distances, binary methylation status, and $\beta$ values,
\item \emph{genomic position}: binary values indicating co-localization with DNA sequence annotations, including promoters, gene body, intergenic region, CGIs, CGI shores and shelves, and SNPs (in site probes); 
\item \emph{DNA sequence properties}: continuous values representing the local recombination rate from HapMap~\cite{TheInternationalHapMapConsortium2003}, GC content from ENCODE~\cite{TheENCODEprojectconsortium2004}, integrated haplotype scores (iHS)~\cite{Voight2006}, and genomic evolutionary rate profiling (GERP) calls~\cite{Dovydov2010}.
\item \emph{CREs}: binary values indicating co-localization with cis-regulatory elements (CREs), including DHS sites (assayed in the GM12878 cell line, the closest match to whole blood) and $79$ specific TFBSs (assayed in the GM12878 cell line)~\cite{TheENCODEprojectconsortium2004}.
\end{itemize}

We used a random forest (RF), which is an ensemble classifier that builds a collection of decision trees and combines the predictions across all of the trees to create a single prediction. The output from the random forest is the proportion of trees in the fitted forest that classify the test sample as a 1. We threshold this output, $\hat{\beta}\in [0,1]$, to $\{0,1\}$ using a cutoff of $0.5$ to find the predicted methylation status. We quantified generalization error for each feature set using a modified version of repeated random subsampling, where the training and test sets are within the same individual (see Materials and Methods).
We used prediction accuracy, specificity, sensitivity, and area under the Receiver Operating Characteristic (ROC) curve (AUC) to evaluate our predictions (see Materials and Methods). 

Using all $97$ features and not restricting neighboring site distance, we achieved an accuracy of $91.6\%$ and an AUC of $0.96$. We considered the role of each subset of features (Table~1). For example, if we only include \emph{genomic position} features,  
the classifier had an accuracy of $78.6\%$ and AUC of $0.84$. 
When we included all classes of features except for \emph{neighbors}, the classifier achieved an accuracy of $85.7\%$ and an AUC of $0.92$, supporting substantial improvement in prediction from considering the methylation status of neighboring CpG sites (t-test $p=4.54\times 10^{-26}$).
However, we also found that the additional features improve prediction substantially over just using the \emph{neighbors} features, which has an accuracy of $90.7\%$ and an AUC of $0.94$ (t-test $p=9.29\times 10^{-13}$).

\begin{table}[h] \scriptsize
\caption{\bf {Performance of methylation status prediction using different prediction models.} AUC: area under ROC curve; MCC: Matthew's Correlation Coefficient; distance: the genomic distance between neighboring CpG sites; gene\_pos: genomic position features including gene region status (promoter, gene body, and intergenic region), CGI status (CGI, CGI shore, CGI shelf, and non-CGI), and proximal SNPs; seq\_property: DNA sequence properties include GC content, recombination rate, conservation score, integrated haplotype scores; CREs include TFBSs and DHS sites.}
\makebox[\linewidth]
{\begin{tabular}{llllllll}\toprule
Feature set & Features & Distance & Accuracy (\%) &  AUC &  Specificity (\%) &  Sensitivity (\%) &  MCC \\\midrule
Gene\_pos & 9 & Arbitrary & 78.6 & 0.84 & 72.6 & 83.5 & 0.57 \\
Gene\_pos + seq\_property & 13 & Arbitrary & 79.4 & 0.86 & 71.8 & 85.6 & 0.58 \\
Gene\_pos + seq\_property + CREs &  93 & Arbitrary & 85.7 & 0.92 & 78.2 & 91.8 & 0.71 \\
\multirow{3}{*}{Neighbor CpG methylation status and distance} & \multirow{3}{*}{4} & Arbitrary & 90.7 & 0.94 & 87.1 & 93.7 & 0.81 \\
 &  & 5 kb & 91.7 & 0.96 & 93.7 & 89.1 & 0.83 \\
 &  & 1 kb & 94.0 & 0.97 & 96.6 & 88.4 & 0.86 \\
\multirow{3}{*}{All features} & \multirow{3}{*}{97} & Arbitrary & 91.6 & 0.96 & 87.9 & 94.6 & 0.83 \\
 &  & 5 kb & 92.5 & 0.97 & 92.8 & 92.1 & 0.85 \\
 &  & 1 kb & 94.3 & 0.98 & 96.0 & 90.8 & 0.87 \\\bottomrule
\label{tab:diffFeaturePredict}
\end{tabular}}
\end{table}

Because the correlation between neighboring CpG sites' methylation levels decays rapidly with distance, we considered restricting the window size for neighboring CpG sites to $1$ kb, $5$ kb, and arbitrary distances on the same chromosome: predictions at CpG sites where one or both of the neighboring CpG sites were greater than this distance were excluded. There were $265,950$ sites ($70\%$) with both upstream and downstream neighboring CpG sites within $5$ kb and $189,735$ sites ($50\%$) with both neighboring sites within $1$ kb. We found that when we restricted prediction to $5$ kb and $1$ kb windows, the classifier using only the \emph{neighbors} features had accuracies of $91.7\%$ and $94.0\%$, and AUCs of $0.96$ and $0.97$, respectively (Table~1). This result is consistent with the high correlation of neighboring CpG sites within $1$ kb. When we used all of the prediction features and restricted neighboring CpG sites to $1$ kb, the classifier achieved $94.3\%$ accuracy and an AUC of $0.98$. Interestingly, while prediction accuracy, AUC, and specificity all improved when restricting neighboring sites to shorter genomic windows, sensitivity, or $TP/(TP+FN)$, was increasingly worse, indicating that the statistical gain in restricting site distance is in eliminating type I errors (false positives) rather than in predicting a higher proportion of methylated sites correctly (true positives; Table 1).

To determine how predictive methylation profiles were across individuals, we quantified the generalization error of our classifier genome-wide across individuals. In particular, we trained our classifier on 10,000 sites from one individual, and predicted CpG sites from the other $99$ individuals for all feature sets. The classifier performance was highly consistent across genomic regions and individuals regardless of the subset of CpG sites in the training set (Figure 6A and Figure S3). We found that the AUC increased when restricting the distance of neighboring CpG sites, repeating earlier trends (t-test between arbitrary distances and $1$ kb: $p < 2.2\times 10^{-16}$).

\begin{figure}[!tpb]
\centerline{\includegraphics[width=6in]{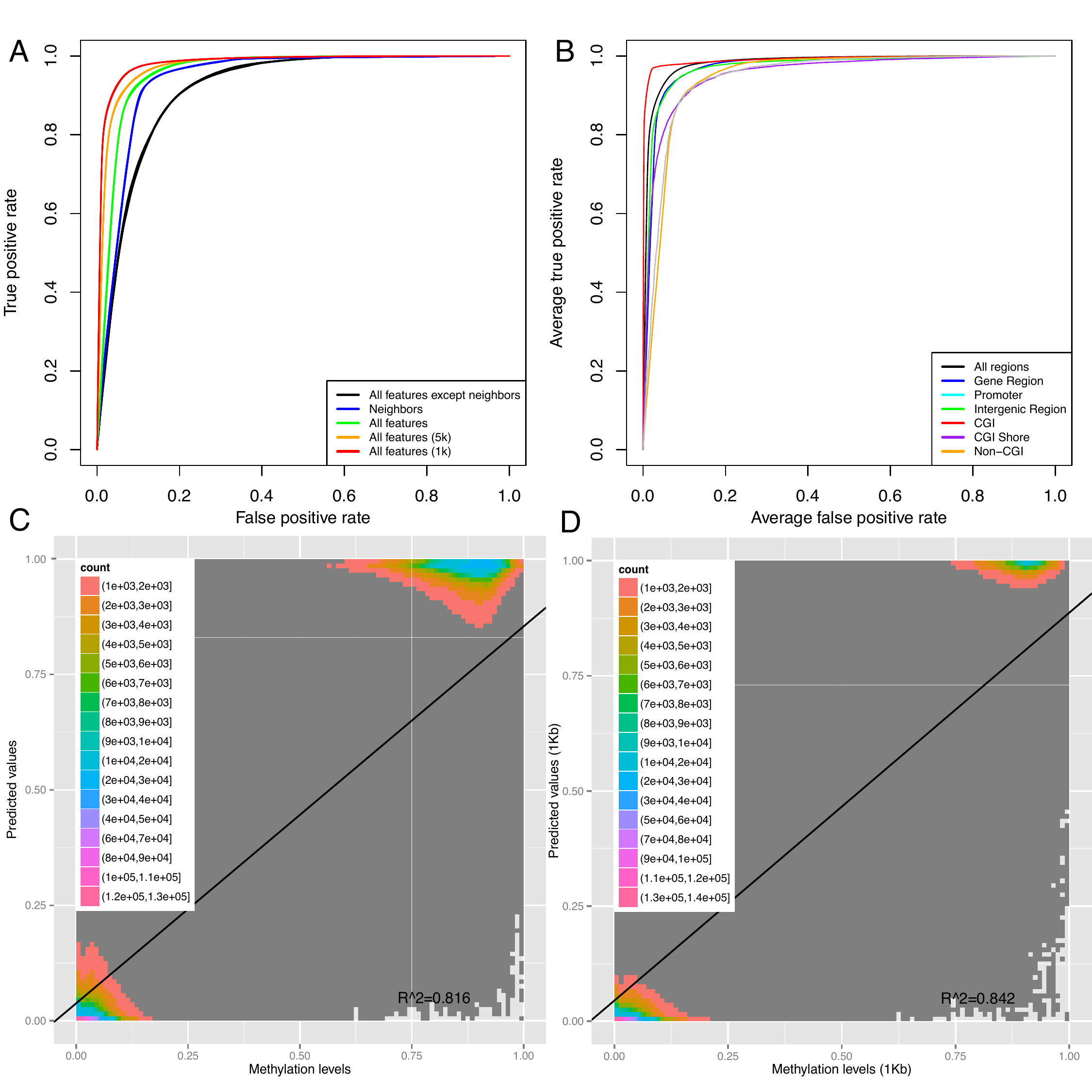}}
\caption{{\bf Prediction performance of methylation status and level prediction.}
       Panel A: ROC curves for prediction of methylation status prediction. The different colors represent different feature combinations used in the classifiers. For each category, the curve was generated by averaging the results of all held out test sets.
       Panel B: ROC curves of region specific methylation status prediction. The different colors represent prediction on CpG sites within different genomic regions.  For each category, the curve was generated by averaging the results of all held out test sets.
       Panel C: 2D histogram of predicted methylation levels versus experimental methylation $\beta$ values. The color represents the density of the distribution at each point. The line is fitted by linear regression.
       Panel D: 2D histogram of predicted methylation levels versus experimental methylation $\beta$ values with restricted to 1 kb neighboring CpG sites. The color represents the density of the distribution at each point.}
\end{figure}

To test the sensitivity of our classifier to the number of CpG sites in the training set, we investigated the prediction performance for different training set sizes. We found that training sets with greater than $1,000$ CpG sites had fairly similar performance (Figure~S4). Throughout these experiments, we used a training set size of $10,000$, in order to strike a balance between large numbers of training samples and computational tractability.


We compared the prediction performance of our random forest classifier with an SVM classifier with a radial basis function (RBF) kernel. We validated by using repeated random resampling, with the same training and test sets for both classifiers. SVM classifiers had an accuracy of $90.6\%$ including all features and neighbors at an arbitrary distance; the accuracy was $92.2\%$ when neighboring sites were restricted to $5$ kb window and $94.5\%$ when restricting to $1$ kb window (Table S5). We found that the random forest classifier had better prediction accuracy than the SVM classifier without restricting neighboring CpG site distance ($p=4.095\times 10^{-14}$; Table~S5). When we restricted neighboring CpG sites to within $1$ kb, random forest and SVM classifiers had similar prediction performance.

\subsection*{Predicting genome-wide methylation levels}

CpG methylation levels $\beta$ in a DNA sample represent the average methylation status across the cells in that sample and will vary continuously between 0 and 1 (Figure S5). 
Since the Illumina 450K array enables precise methylation levels at CpG site resolution in each sample, we used our classifier to predict methylation levels at single CpG site resolution. We compared the prediction probability (${\hat \beta} \in [0,1]$) from our random forest classifier (without thresholding) with methylation levels (${\beta} \in [0,1]$) from the array, and validated this approach using repeated random subsampling to quantify generalization accuracy (see Materials and Methods). Using all $97$ features used in methylation status prediction, but modifying the \emph{neighbors} neighboring CpG site methylation status values to be continuous methylation levels $\beta$, we trained our random forest classifier and evaluated the correlation coefficient ($r$) and root-mean squared error (RMSE) between experimental and predicted methylation values (Table~2; Figure~6C and 6D). We found that the experimental and predicted methylation values had an $r = 0.90$, which increased to $0.92$ and $0.94$ when restricting neighboring CpG sites to within $5$ kb and $1$ kb, respectively. The RMSE between experimental and predicted methylation values for the unrestricted experiment was $0.19$, decreasing to $0.17$ and $0.15$ for the $5$ kb and $1$ kb restrictions. The correlation coefficient and the RMSE indicate good recapitulation of experimental values using predicted methylation values. 

\begin{table}[h]
\caption{\bf {Performance of methylation level predictions using the random forest classifier.} The rows represent distance restrictions on the pairs of neighboring sites (\emph{Arbitrary} for none). R represents the correlation and RMSE represents the root mean squared error of the predicted and actual methylation levels in the CpG sites.}
{\begin{tabular}{lll}\\\toprule 
Distance & R & RMSE \\\midrule
Arbitrary & 0.9036 & 0.1936 \\
5 kb & 0.9198 & 0.1702 \\
1 kb & 0.9356 & 0.1466 \\\bottomrule
\label{tab:LevelPerformance}
\end{tabular}}
\end{table}
\subsection*{Region specific methylation prediction}

Studies of DNA methylation have focused on methylation within CGIs at promoter regions, restricting predictions to CGI regions~\cite{Bock2006, Fang2006, Fan2008, Previti2009, Lu2010, Zhou2012, Zheng2013}; we and others have shown DNA methylation has different patterns in different genomic regions~\cite{Jones2012}. Here we investigated regional DNA methylation prediction for our genome-wide CpG site methods, restricted to CpGs within gene coding regions and CGIs (Table~S6).  

For this experiment, prediction was performed on CpG sites with neighboring sites within $1$ kb distance because of the limited size of CGIs. We found, within CGI regions, predictions of methylation status using our method had an accuracy of $98.3\%$ and an AUC of $0.99$. Methylation level prediction within CGIs achieved a correlation coefficient of $0.94$ and a RMSE of $0.09$. As with the related work on prediction within CGI regions, we believe the improvement in accuracy is due to the limited variability in methylation patterns in these regions; indeed, $90.3\%$ of CpG sites in CGI regions have a $\beta < 0.5$ (Table~S6). 

Conversely, prediction of CpG methylation status within CGI shores had an accuracy of $89.89\%$. This lower accuracy is consistent with observations of robust and drastic change in methylation status across these regions~\cite{Irzarry2009, Doi2009}. Prediction performance within various gene regions was fairly consistent, with $94.9\%$ accuracy for predictions of CpG sites in promoter regions, $93.3\%$ accuracy for predictions within gene body regions (exons and introns),
and $92.9\%$ accuracy within intergenic regions (Figure~6B). This pattern may in part reflect the biased density of CpG sites on the Illumina 450K array. 

\subsection*{Feature importance for methylation prediction}

We evaluated the contributions of each feature to the overall prediction accuracy, as quantified by the Gini index. The \emph{Gini index} measures the decrease in \emph{node impurity}, or the relative entropy of the observed positive and negative examples before and after splitting the training samples on a single feature, of a given feature over all trees in the trained random forest. We computed the Gini index for each of the $97$ features from the fitted RF for predicting methylation status.
We found that upstream and downstream neighboring CpG site methylation status are the most important features (Table~S2, Figure 7A). When we restrict the neighboring CpG sites to be within $5$ kb or $1$ kb, the Gini score of the neighboring site status features increased in relation to the other features, echoing our observation that the non-\emph{neighbor} feature sets are less useful when a CpG site's neighbors are nearby, and thus more informative. In contrast, the Gini score of the genomic distance to the neighboring CpG site feature decreased when the neighboring site distance was restricted, suggesting that neighboring genomic distance an important feature to consider when some neighbors may be more distant and correspondingly less predictive. We found that DHS sites are strongly predictive of an unmethylated CpG site; the DHS site feature has the third most significant Gini index across these experiments. This observation is consistent with a previous study showing that CpG sites in DHS sites tend to be unmethylated~\cite{Tsumagari2013}. 
CGI status is also an important feature, which is unsurprising given that most CpG sites in CGIs are unmethylated. 
GC content, which also ranked highly based on Gini index, may have a substantial contribution to prediction as a proxy for other important features, such as CGI status and CpG density. 
  
\begin{figure}[h]
\centerline{\includegraphics[width=4in]{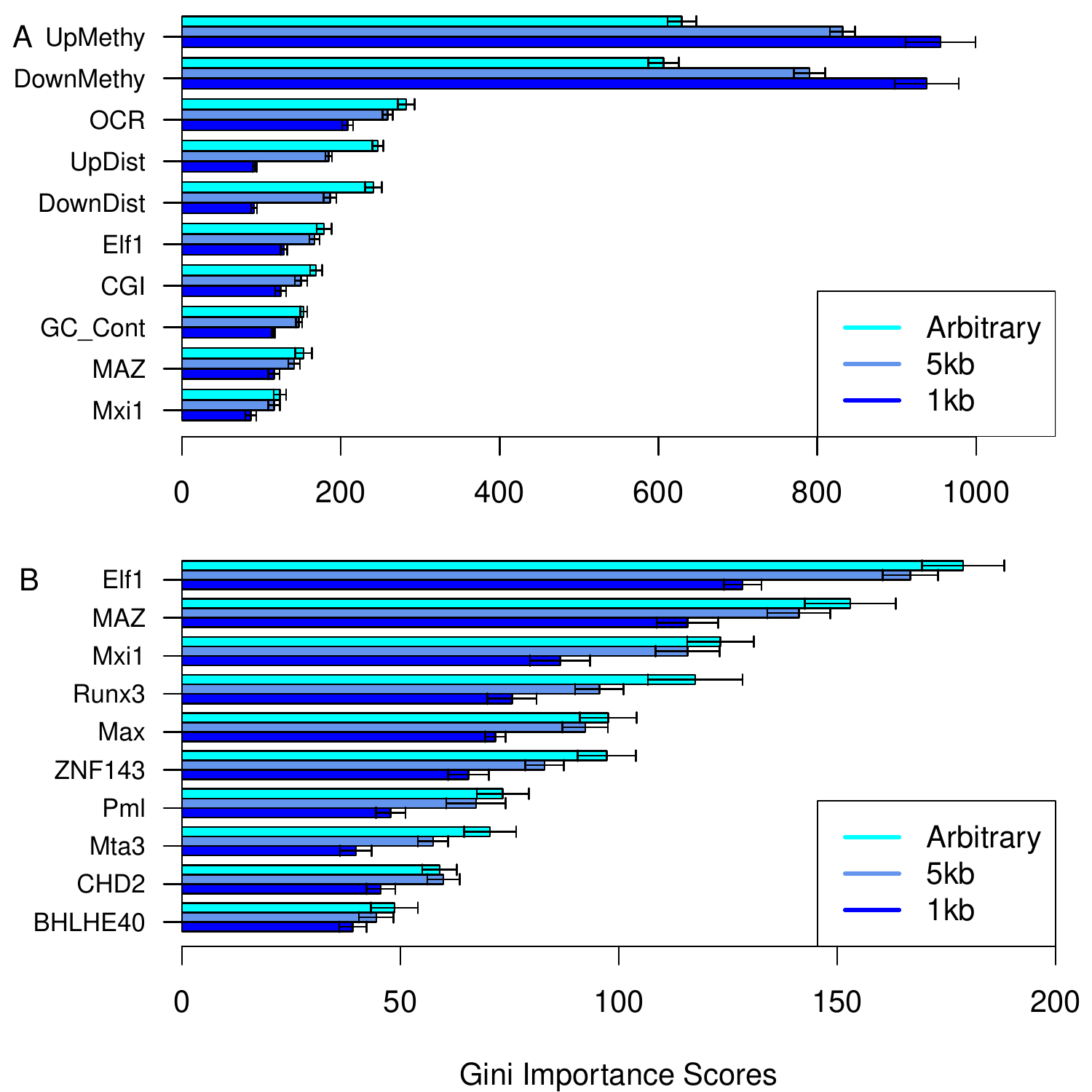}}
\caption{{\bf Ten most important features and TFBSs by Gini score.} The different colors represent Gini scores for features in fitted classifiers with neighboring CpG sites at arbitrary distance, and restricted to $5$ kb and $1$ kb.
      Panel A: Gini scores for the top 10 features.  UpMethy: upstream CpG site methylation status; DownMethy: downstream CpG site methylation status; DHS sites: DNAse I hypersensitive sites; UpDist: Genomic distance with upstream CpG site; DownDist: Genomic distance with downstream CpG site; CGI: CGI status; GC\_Cont: GC content, MAZ, Elf1, Mxi1: TFBSs of MAZ, Elf1 and Mxi1.
      Panel B: Gini scores for the top 10 TFBS.}
\end{figure}

Several TFs were among the most highly ranked features across experiments (Figure~7B), some of which are known to be associated with DNA methylation, including Elf1, Runx3, MAZ, Mxi1, and Max. Indeed, the ETS-related transcription factor (Elf1) has been shown to be overrepresented in methylated regions, associating DNA methylation with hemapotopoiesis in hematopoietic stem cells~\cite{Hogart2012}. 
Runx3 (Runt-related transcription factor 3), a strong tumor suppressor associated with diverse tumor types, has been suggested to be associated with cancer development through regulating global DNA methylation levels~\cite{Chuang2010, Li2002, Kim2005, Lau2006, Sato2006, Weisenberger2006}. Runx3 expression is associated with aberrant DNA methylation in adenocarcinoma cells~\cite{Sato2006}, primary bladder tumor cells~\cite{Kim2005}, and breast cancer cells~\cite{Lau2006}. For another tumor suppressor transcription factor00, Mxi1 (MAX-interacting protein 1), expression levels (specifically, lack of expression) have been reported to be associated with promoter methylation levels and neuroblastic tumorigenesis~\cite{Lazcoz2007}.
It has been suggested that suppression of MAZ (Myc-associated zinc finger protein) may be associated with DNA methyltransferase I, the key factor for {\it de novo} DNA methylation~\cite{Song2001, Song2003}.  
Mxi1 and MAX (Myc-associated factor X) both interact with c-Myc (myelocytomatosis oncogene), a well characterized oncogene, 
which has been shown to be methylation sensitive, meaning that the TF motifs contain CpG sites and, thus, TF binding is sensitive to methylation status at those sites~\cite{Baron2012}. This suggests a potential regulatory relationship between MAX, Mxi1, and DNA methylation that may extend to downstream cancer tumor development.

We found that the feature rankings based on Gini scores differed only slightly across experiments (Figure 7A and 7B), indicating the robustness of the Gini index to quantify contribution to prediction accuracy. We found that the correlation between a binary feature and PC1 is proportional to the Gini index of that feature (Figure 5 and Table S2).
The variation in the Gini index rankings for TFBSs varied more than we expected based on the other features (Figure S6). TFBSs that co-occur with CpG sites more often tend to be more important for prediction, according to the Gini index score. We found that the Gini index of a binary feature has a log linear relationship with the number of co-occurrences of that binary feature with CpG sites in the data set: the more often a CpG site in the training set co-occurred with a TFBS, the higher the Gini index rank of that CpG site. There were several outliers to this trend, including upstream and downstream CpG site methylation status, status of promoter, gene body, and non CGI regions, and three TFBSs (Pol3 (RNA polymerase III), Rxr-$\alpha$ (Retinoid X receptor alpha) and C-fos (a proto-oncogene)). These features were less important than we would predict using the fitted linear regression model of log Gini importance. 
This trend limits the strong conclusions that associate specific TFBSs with DNA methylation biochemically from a high Gini index rank for that TFBS; it may be that there are general relationships between TFBS and CpG sites that we are learning, but a relatively high TFBS frequency in these data will artificially inflate the rank of that TFBS in comparison to the others (Figure~S6). Most CpG sites within TFBSs have low average methylation levels (Table S3). Several TFBSs have disproportionately high average methylation values: for example ZNF274 (Zinc-finger protein 274) and JunD (Jun D proto-oncogene); however, both of these outliers also have a low co-occurrence frequency with CpG sites in these data, suggesting that this finding may be an artifact.

\section*{Discussion}

We characterized genome-wide and region-specific patterns of DNA methylation. These region-specific patterns raise additional questions, including how these observations may resolve or at least suggest causal relationships between methylation and other genomic and epigenomic processes. With single nucleotide polymorphism (SNP) associations with complex traits, it is likely that the genotype drives associated processes rather than the other way around; the causal relationship is established by inductive logic, since it is biologically difficult to perform site-specific mutation. The dynamic nature of CpG site methylation means that no such causal relationship can be established inductively; however, experiments can be designed to establish the impact of changing the methylation status of a CpG site~\cite{Toyota2010, Esteller2000}. Conditional analyses, such as those developed for DNA, may prove to be illuminating for epigenomics~\cite{Yang2012}, but the current data are still murky. For example, does a TFBS containing a CpG site prevent methylation when a transcription factor is actively bound, or does a methylated CpG site in a TFBS prevent a TF from binding to that site?

We built a random forest predictor of DNA methylation levels at CpG site resolution.  In our comparison between random forest classifiers and SVM classifiers, we found that improvements of the random forest classifier include i) better prediction for more sparsely sampled genomic regions, and ii) biological interpretability that comes from the ability to readily extract information about the importance of each feature in prediction. The accuracy results for predictions based on this model are promising, and suggest the possibility of imputing CpG site methylation levels genome-wide in the future. For example, if we assay a set of individuals in an EWAS study on the Illumina 450K array, we may be able to impute the missing genome-wide CpG sites from WGBS assays.  However, we are still far from the prediction accuracies currently expected for SNP imputation for downstream use in GWAS studies. Our cross-sample analysis illustrates that including methylation profiles from other individuals as references, as is done for DNA imputation~\cite{Howie2009}, may improve accuracies substantially. However, because of biological, batch, and environmental effects on DNA methylation, it is possible that precise imputation will require a much larger reference panel relative to DNA imputation. As in GWAS studies, all of these imputation methods will fail to predict rare or unexpected variants~\cite{Howie2012}, which may hold a substantial proportion of association signal for both GWAS and EWAS~\cite{Zhu2011, McClellan2010}. This work raises the additional question, then, of how best to sample CpG sites across the genome given the methylation patterns and the possibility of imputation; for example, it may be sufficient to assay a single CpG site within a CGI and impute the others, given the high correlation between methylation values in CpG sites within the same CGI.


We identified genomic and epigenomic features that were most predictive of methylation status for co-located CpG sites. 
The biological functions of CGI shore and shelf regions, and in particular the impact of methylation in these regions, are mostly unknown; however, it has been shown there is substantial DNA methylation variation in CGI shore regions relative to other regions in the genome, and these alterations may contribute to cancer development and tissue differentiation~\cite{Irzarry2009, Doi2009}. We hope to better characterize the role of CGI shore and shelf regions with respect to enrichment of particular regulatory elements in the future to understand the cellular role of these regions and the specific, curious pattern of methylation found within them.

One particularly important driver of methylation that we do not study carefully here is methylation quantitative trait loci (meQTLs), or genetic drivers of methylation~\cite{Bell2011, Gibbs2010, Zhang2010}.
There is substantial work on the enrichment of meQTLs within SNPs and genetic loci that appear to regulate gene transcription levels (eQTLs), DHS site status (dsQTLs), and others~\cite{Bell2011, Tsumagari2013, Gibbs2010, Degner2012, Pai2012, Gaffney2012}. The characterizations described here lead us to consider identifying QTLs associated with deviations from CRE-specific methylation patterns instead of single CpG sites, as has been done with methylation in CGI shore regions and associations with cancer~\cite{Doi2009}.

\section*{Conclusion}

We investigated genome-wide methylation in 100 individuals profiled using the Illumina 450K array. We identified patterns of correlation in DNA methylation at CpG sites specific to CpG islands, CGI shores, and non-CGIs, quantifying the variability within CGI shore regions and a pattern of correlation across the shelf regions by which correlation increases with distance. We built a random forest classifier to predict methylation as a binary status and as a continuous level at single CpG site accuracy, using as features neighboring CpG site information, genomic position, DNA sequence properties, and cis-regulatory element co-location information. Our method  outperformed state-of-the art methylation classifiers, including our own version of an SVM-based classifier. Our approach quantifies features that are most predictive of CpG status: we found that neighboring CpG site methylation levels, location in a CpG island, and co-localized DHS sites and specific transcription factor binding sites were most predictive of DNA methylation levels. We identify several TFBSs, including Elf1, MAZ, Mxi1, and Runx3, that are highly predictive of methylation levels in whole blood. These predictive features may play a mechanistic role in methylation, either in regulating the methylation of CpG sites or as a downstream partner in modifying the cellular phenotype.

\section*{Materials and methods}

\subsection*{DNA methylation data}

Illumina HumanMethylation450K array data were obtained for 100 unrelated individuals from the TwinsUK cohort~\cite{Moayyeri2012}
All participants in the study provided written informed consent in accordance with local ethics research committees. The 100 individuals were adult unselected volunteers and included 97 female and 3 male individuals (age range 27--78). Whole blood was collected and DNA was extracted using standard protocols. 

Illumina HumanMethylation450K array (Illumina 450K) measured DNA methylation values for more than $482,000$ CpG sites per individual at single-nucleotide resolution. The genomic coverage includes $99\%$ of reference sequence genes, with an average of $17$ CpG sites per gene region distributed across the promoter, 5'UTR, first exon, gene body, and 3'UTR, and $96\%$ of CpG islands~\cite{Rechache2012, Bibikova2011}.

Methylation values for each CpG site are quantified by the term $\beta$, which is the fraction of methylated bead signal over the sum of methylated and unmethylated bead signal:

\begin{equation}
\beta =\frac{\max (Methy,0)}{\max (Methy,0)+\max (Unmethy,0)+\alpha }
\end{equation}

where \emph{Methy} represents the signal intensity of the methylated probe and \emph{Unmethy} represents the signal intensity of the unmethylated probe. The quantity $\beta$ ranges from $0$ (unmethylated) to $1$ (fully methylated).

Data quality control was implemented using R (http://www.r-project.org/) (version 2.15.3). We removed $17,764$ CpG sites for whom the probes mapped to multiple places in the human genome reference sequence. CpG sites with missing values or detection p-values $>0.01$ were excluded. Methylation data from the X and Y chromosomes were excluded, leaving $394,354$ CpG sites from $100$ individuals in downstream analyses. 
The data were controlled for array number, sample position on the array, age, and sex by taking the residual from a fitted linear regression model. The sum of residuals and intercepts of each site was scaled to $[0,1]$ by truncating all sites with values larger than $1$ to $1$ and all sites with values smaller than $0$ to $0$.  
Data quality was assessed to identify sample outliers and batch effects using principal component analysis (PCA)~\cite{Gabriel1990}, no obvious outliers were identified.

\subsection*{Correlation and PCA}

The statistical analyses were implemented using R and Bioconductor (http://www.bioconductor.org/) (version 2.15.3).
Methylation correlations between CpG sites were assessed by the absolute value of Pearson's correlation coefficient and mean square error (MSE):
\begin{equation}
MSE=\frac{\sum_{i=1}^{n}(x_{1i}-x_{2i})^{2}}{n},
\end{equation}
where $x_{1i}$ and $x_{2i}$ represent the methylation values of the two CpG sites being compared, $n$ represents the total number of CpG sites being compared. For \emph{neighboring} CpG sites, pairs of CpG sites assayed on the array that were adjacent in the genome were sampled; the genomic distance between the pairs of CpG sites were within the range $x -200$ bp to $x$ bp, where $x \in \{200, 400, 600,\dots, 6000\}$. The correlation and MSE of a $200$ bp window was not computed, as there were too few CpG sites.  The non-adjacent pair correlation or MSE values are the average absolute value correlation or MSE of $5000$ pairs of CpG sites that were not immediate neighbors with their genomic distances in the same range as for the adjacent CpG sites.

We performed PCA on methylation values of CpG sites by computing the eigenvalues of the covariance matrix of a subsample of CpG sites using the R function {\tt svd}. Among the $378,677$ CpG sites that have complete feature information, $37,868$ sites (every tenth CpG site) were sampled along the genome across all autosomal chromosomes.
Pearson's correlation was calculated between each feature and first ten PCs. PCA analysis was performed by plotting the principal component biplot (scatterplot of first two PCs). The prediction performances were assessed by Receiver Operating Characteristic (ROC) curves and residual sum of squares (RSS). 

\subsection*{Random forest classifier}

We used the {\tt randomForest} package in R for the implementation of the random forest classifier~\cite{RRandForest} (version 4.6-7).
Most of the parameters were kept as default, but {\tt ntree} was set to $1000$ to balance efficiency and accuracy in our high-dimensional data. We found the parameter settings for the random forest classifier (including the number of trees) to be robust to different settings, so we did not estimate parameters in our classifier. The Gini index, which calculates the total decrease of node impurity (e.g., the relative entropy of the class proportions before and after the split) of a feature over all trees, was used to quantify the importance of each feature:

\begin{equation}
I(A)=1-\sum_{k=1}^{c}p_{k}^{2},
\end{equation}
where $k$ represents the class and $p_{k}$ is the proportion of sites belonging to class $k$ in node $A$. 

We compared the performance of the random forest with the support vector machine (SVM) an alternative classifier frequently used in related work~\cite{Bhasin2005, Bock2006, Das2006, Fang2006, Fan2008, Previti2009, Eckhardt2006}. We built an SVM classifier with a Radial Basis Function (RBF) nonlinear kernel, which generally yields more accurate results than the linear kernel that was used in most of the related classifiers~\cite{Hsu2010, Arora2012}. We used the SVM implemented in the {\tt e1071} package in R~\cite{RE1071}. The parameters of the SVM were optimized by 10-fold cross-validation using grid search. The penalty constant $C$ ranged from $2^{-1}, 2^{1},...,2^{9}$ and the parameter $\gamma$ in the kernel function ranged from $2^{-9}, 2^{-7},...,2^{1}$. The parameter combination that had the best performance was used in comparisons with our random forest classifier; specifically, we set $ \gamma = 2^{-7}$ and $C = 2^{3}$ 

\subsection*{Features for prediction}

A comprehensive list of $99$ features were used in prediction (Table S4). The \emph{neighbors} features were obtained from data from the Methylation 450K Array; The \emph{position} features, including gene coding region category, location in CGIs, and SNPs, were obtained from the Methylation 450K Array Annotation file; DNA recombination rate data was downloaded from HapMap (phaseII\_B37, update date Jan\-19\-2011)~\cite{TheInternationalHapMapConsortium2003}; GC content data were downloaded from the raw data used to encode the gc5Base track on hg19 (update date Apr\-24\-2009) from the UCSC Genome Browser (http://hgdownload.cse.ucsc.edu/goldenPath/hg19/gc5Base/)~\cite{Meyer2013}, integrated haplotype scores (iHS scores) were downloaded from the HGDP selection browser iHS data of smoothedAmericas (update date Feb\-12\-2009) (http://hgdp.uchicago.edu/data/iHS/)~\cite{Voight2006}, and GERP constraint scores were downloaded from SidowLab GERP++ tracks on hg19 (http://mendel.stanford.edu/SidowLab/downloads/gerp/)~\cite{Dovydov2010}; \emph{CREs} features: DNAse I hypersensitive sites data were obtained from the DNase-seq data for the GM12878 cell line produced by Crawford Lab at Duke University (UCSC Accession: wgEncodeEH000534, submitted date Mar\-20\-2009) and $79$ specific transcription factor binding sites ChIP-seq data were from the narrow peak files from GM12878 cell line that were available before June 2012 from the ENCODE website~\cite{TheENCODEprojectconsortium2004}.

Neighboring CpG site methylation status was encoded as ``methylated" when the site has a $\beta \geq 0.5$ and ``unmethylated" when $\beta < 0.5$. For continuous features, the feature value is the value of that feature at the genomic location of the CpG site, and for binary features, the feature status indicates whether the CpG site is within that genomic feature or not. DHS sites were encoded as binary variables indicating a CpG site within a DHS site; TFBSs were included as binary variables indicating the presence of a co-localized ChIP-Seq peak; iHS scores, GERP constraint scores and recombination rates were measured in terms of genomic regions; For GC content, we computed the proportion of G and C within a sequence window of $400$ bp, as this feature was shown to be an important predictor in previous study~\cite{Fang2006}.  Among all 99 features, 97 of them (excluding upstream and downstream neighboring CpG sites' $\beta$ values) were used for methylation status predictions, and all excluding upstream and downstream neighboring CpG sites' methylation status were used for methylation level predictions. When limiting prediction to specific regions, e.g., CGI, we excluded those features indicating co-location with that region type. 

\subsection*{Prediction evaluation}

Our methylation predictions were at single CpG site resolution.  
For regional specific methylation prediction, we grouped the CpG sites into either promoter, gene body, intergenic region classes or CGI, CGI shore \& shelf, non-CGI classes according to Methylation 450K array annotation file, which was downloaded from the UCSC genome browser\cite{Kent2002}.

The classifier performance was assessed by a version of repeated random subsampling (RRS) validation. Within a single individual, ten times we pulled $10,000$ random CpG sites from across the genome into the training set, and we tested on all other held-out sites. The prediction performance for a single classifier was calculated by averaging the prediction performance statistics across each of the $10$ trained classifiers. We checked the performance with smaller training set of sizes $100$, $1000$, $2000$, $5000$ and $10,000$ sites in the same evaluation setup. In cross-sample analyses, we set the size of the training set to $10,000$ randomly chosen CpG sites to balance computational performance and accuracy.
We then evaluated the consistency of methylation pattern in different individuals by training the classifier using $10,000$ randomly chosen CpG sites in one individual, and then using the trained classifier to predict all of the CpG sites in the remaining $99$ individuals. 

We quantified the accuracy of the results using the specificity (SP), sensitivity (SE), accuracy (ACC), and the Matthew's Correlation Coefficient (MCC). These values were calculated as follows:

\begin{equation}
SP=\frac{TN}{TN+FP}
\end{equation}
\begin{equation}
SE=\frac{TP}{TP+FN}
\end{equation}
\begin{equation}
ACC=\frac{TP+TN}{TP+FP+TN+FN}
\end{equation}
\begin{equation}
MCC=\frac{TP\times TN-FP\times FN}{\sqrt{(TP+FN)\times (TP+FP)\times(TN+FP)\times (TN+FN)}},
\end{equation}
where TP, TN, FP, FN represent the number of true positives, true negatives, false positives, and false negatives respectively for a particular threshold. Receiver Operating Characteristic (ROC) curves were plotted and we chose a cutoff of $0.5$. The area under the ROC curve (AUC) was calculated; the AUC reflects the overall prediction performance considering both type I (FPs) and type II errors (FNs)~\cite{Bhasin2005, Fogarty2005}. We used the {\tt ROCR} package in R.

To estimate continuous methylation levels ($\beta$), we used the classifier output of prediction probability directly as an estimate of a specific $\beta \in [0,1]$. Prediction accuracy in this regression setting was evaluated using the correlation coefficient and root mean squared error (RMSE).

\begin{equation}
R=\frac{\sum_{i=1}^{n}(x_{i}-x)(y_{i}-y)}{(n-1)\cdot\sigma_{x}\cdot \sigma_{y}}
\end{equation}
\begin{equation}
RMSE=\sqrt{\frac{\sum_{i=1}^{n}(y_{i}-x_{i})^{2}}{n}}
\end{equation}

\noindent where $x_{i},y_{i}$ are the experimental and predicted values, respectively, $x$, $y$ are the means of the experimental and predicted methylation levels, respectively, $\sigma_{x}$, $\sigma_{y}$ are the empirical standard deviations of the experimental and predicted values, respectively.

\bigskip

\section*{Author's contributions}
Conceived the experiments: JTB, BEE. Designed the experiments: WZ, BEE. Performed the experiments: WZ. Analyzed the data: WZ, BEE. Wrote the paper: WZ, JTB, BEE. Responsible for quality control and pre-processing: WZ, JTB, BEE. Contributed valuable data: TDS, PD, JTB.

\section*{Acknowledgements}
  \ifthenelse{\boolean{publ}}{\small}{}
  The authors acknowledge the TwinsUK consortium for providing all methylation data on 100 individuals, Dr. Casey D Brown for providing critical insights, and Dr. Susan Murphy for helpful discussions. BEE was funded through NIH NIGRI R00 HG006265. 
 

\newpage
{\ifthenelse{\boolean{publ}}{\footnotesize}{\small}
 \bibliographystyle{bmc_article}  
  \bibliography{GenomeBiology} }     


\ifthenelse{\boolean{publ}}{\end{multicols}}{}

\end{bmcformat}
\end{document}